\documentclass[conference]{IEEEtran}
\usepackage[sorting=none]{biblatex}
\usepackage{color}
\usepackage{float}
\usepackage{geometry}
\usepackage{graphicx}
\usepackage{hyperref}
\usepackage{subfig}

\title{A Comparison of HDF5, Zarr, and netCDF4 \\in Performing Common I/O Operations}

\author{
\IEEEauthorblockN{Sriniket Ambatipudi}
\IEEEauthorblockA{Dulles High School\\
Houston, TX, USA\\
Email: ambatipudisriniket@gmail.com}
\and
\IEEEauthorblockN{Suren Byna}
\IEEEauthorblockA{Lawrence Berkeley National Laboratory\\
Berkeley, CA, USA\\
Email: sbyna@lbl.gov}
}

\newenvironment{image}[2]
{
\begin{figure}[H]
    \centering
    \subfloat[{\centering Create / Open Times}]{\includegraphics[width=.5\linewidth]{Plots/#1-Create-Open.png}}
    \subfloat[{\centering Read / Write Times}]{\includegraphics[width=.5\linewidth]{Plots/#1-Read-Write.png}}
    \caption{#2}
    \label{fig:#1}
}
{
\end{figure}
}

\newenvironment{config}[3] 
{
\begin{center}
    \begin{tabular}{|p{.375\linewidth}|p{.4375\linewidth}|}
        \hline
        \textbf{Test Name} & {#2 Datasets of} \\
                           & {#3 Elements} \\\hline
        \textbf{File Name} & #1 \\\hline
        \textbf{Number Datasets} & #2 \\\hline
        \textbf{Number Elements} & #3 \\\hline
}
{
    \hline
    \end{tabular}
\end{center}
}

\begin{document}
\maketitle

\begin{abstract}
Scientific data is often stored in files because of the simplicity they provide in managing, transferring, and sharing data. These files are typically structured in a specific arrangement and contain metadata to understand the structure the data is stored in. There are numerous file formats in use in various scientific domains that provide abstractions for storing and retrieving data. With the abundance of file formats aiming to store large amounts of scientific data quickly and easily, a question that arises is, ``Which scientific file format is best suited for a general use case?'' In this study, we compiled a set of benchmarks for common file operations, i.e., create, open, read, write, and close, and used the results of these benchmarks to compare three popular formats: HDF5, netCDF4, and Zarr. 

\emph{Note: This paper is currently a work in progress, and our results are representative for a general-purpose use case in which datasets are small in size and are not using optimizations such as HDF5 or netCDF4 chunking, asynchronous I/O, sub-filing. We welcome any comments or suggestions regarding the benchmark located at \texttt{\url{https://github.com/asriniket/File-Format-Testing}}}.
\end{abstract}

\section*{General Terms / Keywords}
Scientific File Formats, HDF5, netCDF4, Zarr

\section{Introduction}
\label{sec:intro}

With the rapid advancement of experiments, observations, and simulations in recent years, various domains of science are producing enormous amounts of data. For example, the Large Hadron Collider (LHC) experiments at CERN produce 90 petabytes of data per year \cite{CERN-Storage}. NASA’s Climate Data Services (CDS) simulate our planet's weather and climate models from hours to millennia and produce datasets up to petabytes in volume \cite{NASA-CDS}.  

% Brief overview as to why choosing file format is important
Much of the data produced in scientific experiments, observations, and simulations are stored in files with various formats. Scientific file formats offer a medium to store scientific data for long-term processing, which is of great importance to researchers. Each file format has specific characteristics that make it suited for a particular use case, and this makes choosing the appropriate file format an important task. The typical content of scientific files includes data with a structure and metadata describing the structure and the data within. Data in these files are often structured as arrays \cite{Shoshani:book:2009}. The metadata that describes the data often contains the origins of the data, configurations used in generating or collecting the data, and the location of the data in the file format for easy access. As such, scientific file formats will often provide functionality for the storing and retrieval of data and metadata in files. 

% Examples of file formats and their uses in real world scenarios
\par There are numerous file formats in existence, such as HDF5 \cite{HDF5}, netCDF4 \cite{netCDF4}, ROOT \cite{ROOT}, Zarr \cite{Zarr}, and many more \cite{Wikipedia-File-Formats}. Each of these file formats exhibits different performance characteristics and was designed to accomplish a specific task. For example, the High Energy Physics (HEP) community developed the ROOT framework to meet the high-performance requirements for multithreaded read and write operations and support object-oriented programming. As a result of these design specifications, the ROOT framework is used by CERN in its research with the Large Hadron Collider \cite{ROOT-2}. On the other hand, file formats such as netCDF4 or HDF5 are often used in more general use case scenarios because of their self-describing capabilities, which allow the storing of metadata to describe the data within a file. Such self-describing capabilities allow these file formats to be used in a multitude of applications, like how HDF5 is used in astronomy, medicine, physics, and many more fields \cite{HDF5}.

% Give past research that has focused on the testing of file formats
\par Because there are a multitude of file formats that are available to store scientific data, the question of which file format is best suited for a general use case arises. Previous research \cite{Moving-Away-From-HDF5, netCDF4-Performance-Report} has mainly focused on testing the performance characteristics of individual file formats, like how HDF5 was tested for its performance in reading a subset of a large array \cite{Moving-Away-From-HDF5} or how netCDF4's performance characteristics were analyzed \cite{netCDF4-Performance-Report}. The first test revealed that when working with an HDF5 file in Python, the fastest way to read data is to memory map the file with NumPy, bypassing the HDF5 Python API (h5py). Memory mapping involves mapping a file's contents into memory, and this means that data within a file can be accessed if the location of the data in terms of an offset is known. The hierarchical structure of file formats such as HDF5 means that accessing the data within is relatively simple, as the file's metadata stores the location of individual datasets. This test was useful in analyzing the shortcomings of HDF5 in a particular use case, allowing potential users to reconsider whether the HDF5 file format would be best suited for their use case. In the second experiment, conducted by the HDF Group, the netCDF4 file format was tested for its performance characteristics and then compared to its predecessor, netCDF3. The results of this experiment showed that netCDF4 generally had slower write speeds than netCDF3, but it had faster read speeds due to netCDF4's use of the HDF5 library internally.

% Introduce the main problem that there is no test to compare the performance requirements of a file format.
\par This type of testing is useful for analyzing the performance characteristics of one file format and its shortcomings in specific use cases, but when the performance characteristics of one file format must be compared to the performance characteristics of other file formats, a benchmark offers itself as a viable option because it allows for the objective measurement of the speed at which each file format is able to perform a specific task. Such a benchmark has the potential to be a valuable asset to researchers, as it allows them to choose the file format that is not only suited for their particular use case but also performs the best in comparison to its alternatives. This allows the researchers to have ease of access when storing and modifying data at fast speeds, allowing more time and effort to be put elsewhere in their project. In this work, we developed a benchmark to compare the read and write speeds of three multipurpose scientific file formats (HDF5, netCDF4, and Zarr). This benchmark writes randomized data to a specified number of datasets within a file and measures the time taken to write the data to each dataset and the time taken to read the contents of each dataset, allowing objective comparisons to be drawn between the three file formats' performance in different operations.

In the remainder of the paper, we first provide a brief background in \S\ref{sec:bg} to the three file formats we used in our evaluation in this study. In \S\ref{sec:bench}, we describe the read and write benchmarks we used in the evaluation. In \S\ref{sec:eval}, we provide details of the system we used for comparing and evaluating the performance of the three file formats under different workloads. 

\section{Background}
\label{sec:bg}

\subsection{HDF5}
\label{sec:bg:hdf5}

HDF5, or the Hierarchical Data Format 5, is a file format designed to store a large amount of data in an organized manner. Typically characterized by a \texttt{.hdf5} or \texttt{.h5} file extension, this file format stores data in a manner very similar to that of a file system. This file format's primary data models are groups and datasets. Groups are the overarching structure, and they can hold other groups or datasets. Datasets store raw data values of a specified data type and are usually stored within groups \cite{About-HDF5}. A feature of HDF5 is that it is able to store data consisting of different data types within the same file \cite{Introduction-To-HDF5}. As mentioned earlier, this file format is self-describing, meaning that all the groups and datasets within the file format contain metadata describing their contents. This allows for the data within the file to be mapped in memory, provided the API supports it. Generally speaking, users use the HDF5 API to issue commands to a lower-level driver, which is in charge of accessing the file and performing the requested operations \cite{How-HDF5-Works}. Because the file format is open source, there has been widespread API support across most modern languages (Python, C++, and Java).

\subsection{netCDF4}
\label{sec:bg:netcdf4}

netCDF4 is a file format that is designed to store array-oriented data and is characterized by a \texttt{.netc} file extension. It stores data in a manner similar to HDF5, with groups serving as the overarching data structure. Within a group, there can be other groups or variables. Variables are akin to HDF5 datasets. Unlike HDF5 datasets, netCDF4 variables cannot be resized once they are created \cite{netCDF4-Python}. To circumvent this, variables can be declared with an unlimited size in a specified dimension. Similar to HDF5, netCDF4 is a self-describing file format, and this means that groups and variables both contain metadata describing their contents. Unlike its predecessor, netCDF3, netCDF4 uses HDF5 as its backend, allowing it to achieve faster read times \cite{netCDF4-Performance-Report}.

\subsection{Zarr}
\label{sec:bg:zarr}

Zarr is a file format that is designed to store large arrays of data and is characterized by a \texttt{.zarr} file extension. Because it is based on NumPy, it is geared mainly towards Python users. Similar to both HDF5 and netCDF4, Zarr is also a hierarchical, self-describing file format that has groups as the overarching file structure. Each group contains datasets, which are representative of multidimensional arrays of a homogeneous data type. Furthermore, the API for this file format was designed to be similar to h5py (HDF5's Python API), and as a result, it includes functions based on h5py's functions, namely the group creation function \cite{Zarr-2}. One advantage to using Zarr is that it provides multiple options to store data by allowing a user to store a file in memory, in the file system, or in other storage systems with a similar interface to the first two options \cite{Zarr-2}.

\section{Benchmarks}
\label{sec:bench}

% Explain the high-level overview in the benchmark and its significance
As a benchmark is being used to compare the performance of the file formats, the benchmark must only test features of the file format that are supported by all the file formats being tested. To accomplish this task, we programmed our benchmark in Python, and this means we will rely on the Python APIs for each file format being tested to perform the requested operations. Our benchmark compares the time taken to create a dataset, write data to a dataset, and finally open that dataset at a later time and read its contents. This can be categorized into two main types of operations—the writing operation and the reading operation. Both are very important features to test in a file format, as the end goal of a file format is to store data for long-term processing. The faster write and read times are not only indicative of better performance characteristics but also have a tangible effect on the improvement of an end-user's workflow, as less time would be spent performing operations that are not directly relevant to the task at hand.

% Introduce configuration file and transition to each operation.
To allow for greater flexibility when benchmarking the file formats, we added a configuration system in which the user is able to specify the testing parameters such as the number of datasets to create within the file and the dimensions of the array that will be written to each dataset by editing a \texttt{.yaml} configuration file. After the benchmark is done, the program then stores the times taken across multiple trials in a \texttt{.csv} file and plots the data in the \texttt{.csv} file with \texttt{matplotlib.pyplot} to allow a user to make a definitive comparison between the file formats being tested. Below, the main operations, the write operation and the read operation, will be discussed in-depth. 

\subsection{Write Benchmark}
\label{sec:bench:writebench}

The write operation is the first operation to be tested in the benchmark. It creates files with the filename as specified in the configuration file and extensions \texttt{.hdf5} for HDF5 files, \texttt{.netc} for netCDF4 files, and \texttt{.zarr} for Zarr files. The file is placed inside a folder named \texttt{Files/}, to help reduce clutter in the working directory. Taking information from the configuration file, a sample data array is generated with dimensions and length as specified. This sample data array consists of randomly-generated 32-bit floats. Then, the program creates a dataset within the file and writes the sample data array to the dataset. This process of generating a sample data array, creating a dataset, and populating it with the values from the sample data array is repeated until the benchmark has created the number of datasets as specified by the configuration file. After the file is populated with data, the benchmark copies the file to a directory named \texttt{Files Read/} and renames the file to avoid any caching effects that may interfere with the read times. There are numerous ways to mitigate such caching effects, such as waiting for an extended period of time, but simply moving the file to another directory and renaming the file is the quickest and easiest way to mitigate the effects of caching in interfering with the times taken to read from the file. The time taken to create all the datasets and populate them with data is divided by the number of datasets to find the average time taken to create and populate one dataset. Both of these times are then returned to the main program, where they are written to the \texttt{.csv} output file.

\subsection{Read Benchmark}
\label{sec:bench:readbench}

The benchmark now opens the copied file in the \texttt{Files Read/} directory and begins testing the read operations of the three file formats. This operation consists of opening each dataset within the file and printing its contents to the standard output. The time taken to open all the datasets and the time taken to read from all the datasets are once again divided by the number of datasets within the file to find the average time taken to open and read one dataset. Both of these times are then returned to the main program, where they are also written to the \texttt{.csv} output file.

\par This process of running the write operation benchmark and the read operation benchmark is then repeated multiple times in order to ensure the consistency of the data gathered. To avoid filling up the disk with generated test files, the \texttt{Files/} and \texttt{Files Read/} directories are deleted between trials. Finally, the data from the \texttt{.csv} file are averaged out with \texttt{pandas} and plotted with \texttt{matplotlib.pyplot} to allow for visualizing a comparison between the tested file formats in a given operation.

\section{Performance Evaluation}
\label{sec:eval}

\subsection{Experimental setup}
\label{sec:eval:expsetup}

The three file formats were tested on a computer running Ubuntu 18.04.5 with an Intel(R) Xeon(R) Silver 4215R CPU, 196 Gigabytes of RAM, and 960 Gigabytes of solid-state storage provided by a Micron 5200 Series SSD. The version of \texttt{h5py} used to test the HDF5 file format was \texttt{3.6.0}. The version of \texttt{netCDF4} used to test the netCDF4 file format was \texttt{1.5.8}. The version of \texttt{zarr} used to test the Zarr file format was \texttt{2.11.0}.

The benchmark parameters that were used in each run of the test can be found in the tables to the right. Note that the \texttt{Test Name} parameter is automatically generated by the benchmark and is used to create the generated plot's title.

\begin{config}{2048-Vector}{2048}{[128]}\end{config}
\begin{config}{2048-Matrix}{2048}{[128, 128]}\end{config}
\begin{config}{2048-Tensor}{2048}{[128, 128, 128]}\end{config}
\begin{config}{2048-Datasets}{2048}{[256]}\end{config}
\begin{config}{4096-Datasets}{4096}{[256]}\end{config}
\begin{config}{8192-Datasets}{8192}{[256]}\end{config}

\subsection{Data}
\label{sec:eval:data}

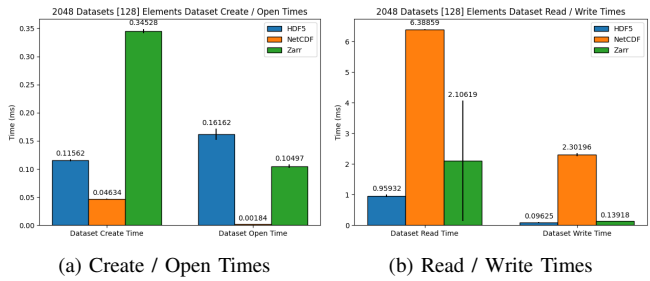
\begin{image}{2048-128}{2048 Datasets of [128] Elements}\end{image}
\begin{image}{2048-128-128}{2048 Datasets of [128, 128] Elements}\end{image}
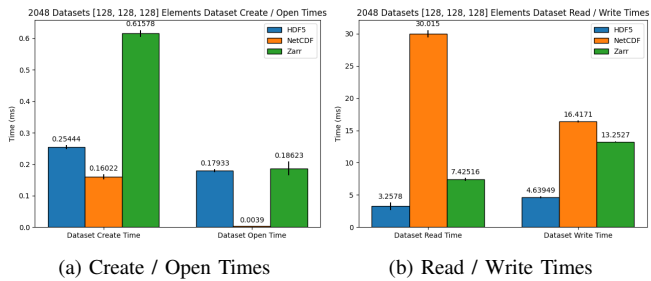
\begin{image}{2048-128-128-128}{2048 Datasets of [128, 128, 128] Elements}\end{image}
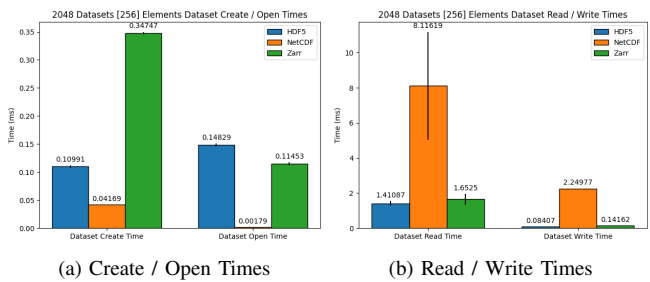
\begin{image}{2048-256}{2048 Datasets of [256] Elements}\end{image}
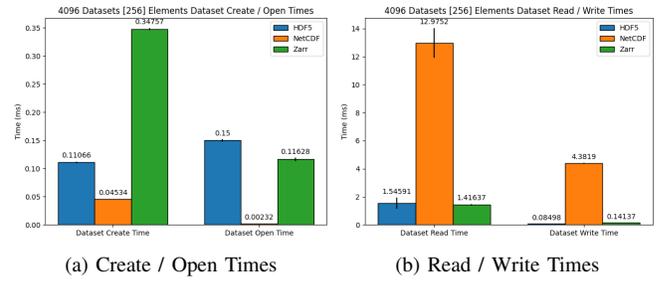
\begin{image}{4096-256}{4096 Datasets of [256] Elements}\end{image}
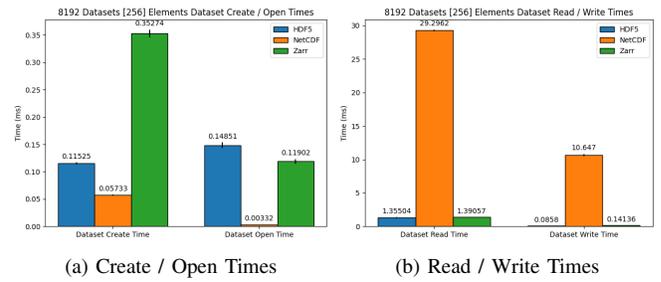
\begin{image}{8192-256}{8192 Datasets of [256] Elements}\end{image}

\subsection{Discussion}
\label{sec:eval:discuss}

Figure \ref{fig:2048-128} shows the results when 2,048 datasets are created and populated with a one-dimensional array containing 128 32-bit floats. This graph shows that the time taken to create and open a dataset in netCDF4 is much faster than that of HDF5's or Zarr's. In comparison to Zarr, HDF5 takes less time to create a dataset, but it takes slightly more time to open a dataset. When it comes to writing to datasets or reading from datasets, HDF5 and Zarr share very similar times in both operations, with netCDF4 trailing by a large margin.

Figure \ref{fig:2048-128-128} shows the results when 2,048 datasets are created and populated with a two-dimensional array containing 128 elements in each dimension. The results from this test are almost identical to the results from the previous test, both in terms of the trend and the time taken to complete each operation.

Figure \ref{fig:2048-128-128-128} shows the results when 2,048 datasets are created and populated with a three-dimensional array containing 128 elements in each dimension. The results from this test follow the same trend as the past two tests, but the times taken to complete each operation are almost double the times taken in the past two tests.

These past three bar graphs show the tests in which the number of datasets is held constant while increasing the number of dimensions in the data array, but the next three bar graphs will involve increasing the number of datasets while holding the size of the data array constant in order to measure the effect of increasing the number of datasets on file-format performance.

Figure \ref{fig:2048-256} shows the results when 2,048 datasets are created and populated with a one-dimensional array containing 256 elements. The results from this test mirror those from Figure \ref{fig:2048-128}, and this is to be expected as the number of datasets in both tests is the same, with the size of each dataset varying slightly.

Figure \ref{fig:4096-256} shows the results when 4,096 datasets are created and populated with a one-dimensional array containing 256 elements, and Figure \ref{fig:8192-256} shows the results when 8,192 datasets are created and populated with a one-dimensional array containing 256 elements. Both graphs are almost identical to Figure \ref{fig:2048-256}, meaning that the number of datasets most likely has no impact on the average time taken to perform the operations requested.

\subsection{Write Benchmark Discussion}
\label{sec:eval:writediscuss}

The results of this benchmark show that a general trend is that when creating a dataset, netCDF4 takes the least time and is followed by HDF5, which is followed by Zarr. 

When actually writing data to a file, HDF5 takes the least time to write data to a dataset and is followed by Zarr, which is followed by netCDF4—taking on average more than double the time of HDF5.

\subsection{Read Benchmark Discussion}
\label{sec:eval:readdiscuss}

The read benchmarks show results similar to those from the write benchmark. netCDF4 takes the least time to open a dataset and is followed by Zarr, which is followed by HDF5.

When reading the data by printing the dataset values to the standard output, HDF5 takes the least time to read a dataset and is followed by Zarr, which is followed by netCDF4.

\section{Conclusions}
\label{sec:conclusions}
In this paper, we demonstrated a method in which the performance of a file format can be compared to that of another file format through the running of a benchmark that tests performance in operations like create, open, read, write, and close. This paper focused specifically on benchmarking three file formats: HDF5, netCDF4, and Zarr, as these three file formats are considered to be general-purpose scientific file formats due to their storing of various types of data in a hierarchical manner, similar to a file system. 

\par The benchmark was conducted in Python due to the language's widespread use in numerous scientific applications, and as such, the Python API for each file format was tested. To determine the performance of a file format, the time taken to create a dataset, write data to the dataset, open the dataset once the file is closed, and read data from the dataset was measured and plotted in a bar graph. The results of the benchmark show that HDF5 is fastest in reading or writing to a dataset, netCDF4 is fastest in creating or opening a dataset, and Zarr generally trails right behind HDF5 in performance.

\par Future work for this benchmark would include: expanding support to other programming languages, as this would reveal any potential bottlenecks within the language-specific API for a file format; testing more file formats in order to better determine which file format is the fastest; and testing more aspects of a file format, which may include testing performance in specific scenarios (i.e., reading a small subset of a dataset, overwriting a dataset). The code for the benchmark can be found here: \texttt{\url{https://github.com/asriniket/File-Format-Testing}}. We are evaluating further the overheads of using the Python API on the observed performance as well as the impact of caching. Considering the small size of the data, the observed results may have been impacted by caching. This caching effect will typically be reduced when the data sizes are in gigabytes (GB). We also note that many applications work with smaller amounts of data that we used in this study. We encourage readers to try out the benchmarks provided in the GitHub repository and contribute any optimizations.

\section{Acknowledgments}
This effort was supported in part by the U.S. Department of Energy (DOE), Office of Science, Office of Advanced Scientific Computing Research (ASCR) under contract number DE-AC02-05CH11231 with LBNL. 

\printbibliography

@online{About-HDF5,
author = {Leah A. Wasser},
title = {Hierarchical Data Formats - What is HDF5?}, 
url = {https://www.neonscience.org/resources/learning-hub/tutorials/about-hdf5},
}

@online{CERN-Storage,
author = {CERN},
title = {CERN - Storage}, 
url = {https://home.cern/science/computing/storage},
}

@online{HDF5,
author = {HDFGroup},
title = {The HDF5® Library \& File Format - The HDF Group}, 
url = {https://www.hdfgroup.org/solutions/hdf5/},
}

@online{How-HDF5-Works,
author = {HDFGroup},
title = {Chapter 3: The HDF5 File},
url = {https://support.hdfgroup.org/HDF5/doc/UG/FmSource/08_TheFile_favicon_test.html},
}

@online{Introduction-To-HDF5,
author = {HDFGroup},
title = {Introduction to HDF5}, 
url = {https://portal.hdfgroup.org/display/HDF5/Introduction+to+HDF5},
}

@online{Moving-Away-From-HDF5,
author = {Cyrille Rossant},
title = {Moving away from HDF5}, 
url = {https://cyrille.rossant.net/moving-away-hdf5/},
}

@online{NASA-CDS,
author = {NASA},
title = {NASA’s Climate Data Services (CDS)}, 
url = {https://www.nccs.nasa.gov/services/climate-data-services},
}

@online{netCDF4,
author = {UCAR/Unidata},
title = {Unidata | NetCDF}, 
url = {https://www.unidata.ucar.edu/software/netcdf/},
}

@online{netCDF4-Performance-Report,
author = {Choonghwan Lee and MuQun Yang and Ruth Aydt},
title = {NetCDF-4 Performance Report}, 
url = {https://support.hdfgroup.org/pubs/papers/2008-06_netcdf4_perf_report.pdf},
}

@online{netCDF4-Python,
author = {UCAR/Unidata},
title = {NetCDF4 API Documentation}, 
url = {https://unidata.github.io/netcdf4-python/},
}

@online{ROOT,
author = {CERN},
title = {ROOT: analyzing petabytes of data, scientifically.}, 
url = {https://root.cern/},
}

@online{ROOT-2,
author = {Barbara Warmbein},
title = {Big data takes ROOT}, 
url = {https://home.cern/news/news/computing/big-data-takes-root},
}

@article{Shoshani:book:2009,
author = {Shoshani, Arie and Rotem, Doron},
year = {2009},
month = {12},
pages = {},
title = {Scientific data management. Challenges, technology, and development},
isbn = {9780429189272},
journal = {Scientific Data Management: Challenges, Technology, and Deployment},
doi = {10.1201/9781420069815},
}

@online{Wikipedia-File-Formats,
author = {Wikipedia},
title = {List of file formats}, 
url = {https://en.wikipedia.org/wiki/List_of_file_formats#Scientific_data_(data_exchange)},
}

@online{Zarr,
author = {Zarr Developers},
title = {Zarr}, 
url = {https://zarr.readthedocs.io/en/stable/},
}

@online{Zarr-2,
author = {Zarr Developers},
title = {Zarr}, 
url = {https://zarr.readthedocs.io/en/stable/tutorial.html#groups/},
}

\end{document}